\documentclass[useAMS,usenatbib]{mn2e}
\usepackage{amsmath,psfig,epsfig,graphics}

\def\aap{A\&A}

\def\apj{ApJ}

\def\apjl{ApJ}
\def\mnras{MNRAS}
\def\araa{ARA\&A}

\def\ssr{Space Sci. Rev. }
\def\pasj{PASJ}

\def\lesssim{\mathrel{\hbox{\rlap{\hbox{\lower4pt\hbox{$\sim$}}}\hbox{$<$}}}}
\def\gesssim{\mathrel{\hbox{\rlap{\hbox{\lower4pt\hbox{$\sim$}}}\hbox{$>$}}}}

\def\lesssim{\mathrel{\hbox{\rlap{\hbox{\lower4pt\hbox{$\sim$}}}\hbox{$<$}}}}
\def\gesssim{\mathrel{\hbox{\rlap{\hbox{\lower4pt\hbox{$\sim$}}}\hbox{$>$}}}}

\begin{document}

\author[Morandi \& Cui]
{Andrea Morandi${}^1$\thanks{E-mail: amorandi@purdue.edu}, Wei Cui${}^1$\\
$^{1}$ Department of Physics, Purdue University, West Lafayette, IN 47907, USA\\
}

\title[Measuring the gas clumping in Abell~133]
{Measuring the gas clumping in Abell~133}
\maketitle

\begin{abstract}
This paper continues a series in which we developed a non-parametric method to measure inhomogeneities in the gas distribution from X-ray observations of galaxy clusters. In this work, we apply our method to \emph{Chandra} X-ray observations of Abell~133 and present the determination of the gas clumping factor from X-ray cluster data. We find that the gas clumping factor in Abell~133 increases with radius and reaches $\sim 2-3$ at $0.9\,R_{200}$. This is in good agreement with the predictions of hydrodynamical simulations and our previous determination. We then observe a general trend of steepening in the radial profiles of the clumping-corrected gas density beyond $0.3\, R_{200}$, with a logarithmic slope of $\sim 2.6$ at $0.9\,R_{200}$. The observed density profiles appear to be flatter compared to simulations, but in agreement with previous observational findings. In addition, we observe that the measured temperature decreases steadily with radius toward the outskirts of A133, while the entropy increases monotonically with radius, gently flattening in the outer volumes. With respect to theoretical predictions from pure gravitational collapse, the results presented here point to an entropy excess in the central regions, which extends out to large radii. These results suggest that gas inhomogeneities should be treated properly when interpreting X-ray measurements in the envelope of galaxy clusters. We finally discuss how the brightness distribution keeps a record of the large-scale structures formation scenario, providing a snapshot of the 'melting pot' in the virialization region. 
\end{abstract}

\begin{keywords}
cosmology: observations -- cosmology: large-scale structure of Universe --galaxies: clusters: general -- X-rays: galaxies: clusters
\end{keywords}

\section{Introduction}\label{intro}
Galaxy clusters represent the largest virialized structures in the present universe. They form in the densest knots of the cosmic web at the intersection of the filaments, from which they continuously accrete material in the form of dark matter, gas and galaxies. Galaxy clusters, by virtue of their position at the high end of the cosmic mass power spectrum, are useful tracers of cosmic evolution and unique probes of mass distribution on large scales. First, accurate three-dimensional mass measurements of clusters can be used to determine the amount of structure in the universe on scales of $10^{14}-10^{15}M_{\odot}$ and to derive constraints on the cosmological parameters by means of comparison with models of cluster mass distribution. Second, clusters are essentially closed boxes that retain all their gaseous matter, which contains a wealth of information about the physical processes associated with galaxy formation, including feedback from supernovae, AGN, star formation, or galactic winds, and traces out the thermal history of these processes. These complex physical properties are only partially understood and captured with modern hydrodynamical simulations; nevertheless, most of these physical processes can be neglected in the outskirts of galaxy clusters. Indeed, the physics of the ICM in the outer volumes is dominated by the gravity-driven collisionless dynamics of DM and hydrodynamics of the gas. Therefore, the exploration of the virialization region of galaxy clusters ($r\gesssim  R_{500}$\footnote{$R_{500}(R_{200})$ is the radius within which the mean total density is 500(200) times the critical density of the Universe, with $R_{500}\simeq 0.6\, R_{200}$. $R_{200}$ can be regarded as the virial radius of the cluster.}) has recently become very important in cluster cosmology, the underlying physics of the ICM being relatively simple, and hence a comparison with theoretical predictions more tenable. The outskirts of galaxy clusters present also an opportunity to provide constraints on the processes that rule the formation of large-scale structures, providing a snapshot of the 'melting pot' in the virialization region, where infalling cosmic baryons partially convert their kinetic energy into the thermal energy of the hot gas, generating clumpy and turbulent ICM. Nevertheless, cluster outskirts are potentially more susceptible than the cluster inner volumes to departures from virialization, turbulence and bulk flows that result from structure formation processes, which could be a significant source of systematic uncertainties in X-ray measurements of the ICM profiles as well as global cluster properties \citep{roncarelli2013}. The origin of these ICM inhomogeneities must be sought among non-spherical gas distributions, infalling gas lumps, the presence of bubbles of relativistic plasma and turbulent gas motions, imperfect mixing of the gas with different entropies displaced by gas motions, shocks and hierarchical accretion of subhaloes \citep{zhuravleva2013}. This, in turn, can limit the use of galaxy clusters as high-precision cosmological probes. Gas density clumping can also appreciably bias the hydrostatic mass \citep{lau2009}, the integrated Compton parameter from X-ray observations \citep{khedekar2013} and the gas fraction \citep{battaglia2012}. 

Hence, a deeper understanding of the state of the intracluster gas in cluster outskirts is required, albeit difficult because of the very low surface brightness of the X-ray signal, the inhomogeneities of the gas associated with clumps, asymmetries and accretion patterns. Studies of the outskirts have just become feasible with current instrumentation. Thanks to its relatively low particle background compared to X-ray satellites like {\emph Chandra} and {\emph XMM-Newton}, {\emph Suzaku} observations have extended X-ray measurements of the ICM profile out to and beyond the virial radius for several clusters, measuring the physical properties in the cluster outskirts. Initial results pointed to a shallow density/entropy profile in cluster outer volumes \citep[e.g.][]{simionescu2011,george2009,bautz2009}. 

\cite{simionescu2011} found that the entropy profile significantly flattens and the baryon fraction exceeds the cosmic mean at large radii for Perseus, calling for a clumpy distribution of the gas. They argue that clumpiness becomes significant at large radii in order to justify the observables. This conclusion has been strengthened by \cite{walker2013}, who inferred on a sample of clusters the clumpiness which is required to bring the entropy level into agreement with the self-similar predictions in the outskirts. Using hydrodynamical cluster simulations, \cite{nagai2011} also pointed out that gas clumping introduces significant biases in X-ray measurements of the ICM profiles in the outskirts, leading to an overestimate of the observed gas density and flattening of the entropy profile. At $R_{200}$, they estimated a clumping factor bracketed in the range $\sim 1.3-2$. Their results thus suggest that gas inhomogeneities should be treated properly when interpreting X-ray measurements in the envelope of galaxy clusters.

Observations of more clusters and better understanding of systematic uncertainties are required before making robust conclusions. Independent measurements with \emph{Chandra} can shed light on systematic uncertainties in the current measurements \citep[e.g.,][]{ettori2011,moretti2011}: indeed, despite its higher background, Chandra provides a superior angular resolution to images (the point-spread has a 0.5 arcsec full width at half-maximum) which is fundamental in order to: i) remove emission from unrelated sources; ii) to constrain the emission of clusters to the virial radius, especially for higher-redshift cool-core clusters for which there is negligible contribution from the bright cluster core to the emission in the outer volumes, and from secondary scatter by sources outside the field of view.

In our previous work \citep{morandi2013}, we developed a robust non-parametric approach to derive the gas clumpiness in galaxy clusters from X-ray observations. The present paper represents a follow-up to of our previous efforts on studying cluster inhomogeneities out to $R_{200}$. In particular, we apply our method to \emph{Chandra} observations of Abell~133 (A133), which serves as a pilot project for the proposed investigations. A133 is a luminous cool-core galaxy cluster at $z=0.0566$, and it an optimal cluster for this work as it is characterized by ultra-deep exposure of $2.4$~Msec (\emph{Chandra} XVP observations of A133 by Vikhlinin et al., in prep), with several pointings encompassing $R_{200}$ of this cluster. Given that such ultra-deep observations are limited to a very small number of systems, this cluster represents a unique opportunity to study the gas inhomogeneities out to large radii, whose study requires extremely sensitive observations.

The paper is organized as follows. In \S\ref{tecn} we briefly outline the method. In \S\ref{dataan} we summarize the most relevant aspects of the X-ray data reduction and analysis. In \S\ref{concl343} we discuss how the brightness distribution keeps a record of the large-scale structures formation scenario, while \S\ref{conclusion33} is devoted to the conclusions. Throughout this work we assume the flat $\Lambda$CDM model, with matter density parameter $\Omega_{m}=0.3$, cosmological constant density parameter $\Omega_\Lambda=0.7$, and Hubble constant $H_{0}=100h \,{\rm km\; s^{-1}\; Mpc^{-1}}$ where $h=0.7$. Unless otherwise stated, we report the errors at the 68.3\% confidence level.

\section{Non-parametric method for deprojecting X-ray data}\label{tecn}
We briefly outline the methodology used to infer gas density and gas clumping factor in X-ray galaxy clusters. Additional details can be found in \cite{morandi2013}. 

X-ray photons are emitted from the hot ICM primarily through the scattering of electrons off of ions via the thermal bremsstrahlung process. The X-ray surface brightness is then given by:
\begin{equation}
S_X = \frac{1}{4 \pi (1+z)^4}  \int n_{\rm e} n_{\rm p} \Lambda(T,Z) \, dl   \propto C(r) \langle \rho_{\rm gas}(r) \rangle^2,
\label{1.em.x.eq22}
\end{equation}
where $\Lambda(T,Z)$ is the cooling function, $T$ and $Z$ are the three-dimensional gas temperature and metallicity, $n_{\rm e}$ is the electron density, and $l$ is distance along the line of sight. $C$ is the clumping factor given by 
\begin{equation}
C\equiv\frac{\left < n_{\rm e} ^2\right > }{\,\left < n_{\rm e} \right >^2} =1+ \frac{\sigma_{n_e,{\rm intr}}^2}{\left<n_e\right >^2} \ge 1.
\label{clump}
\end{equation}
where $\sigma_{n_e,{\rm intr}}$ is the scatter in the gas density distribution. Note that $C=1$ if the ICM is not clumpy (i.e. a single phase medium characterized by a single temperature and gas density within each radial bin). In the X-ray cluster analyses, it is commonly assumed that $C=1$, and the three dimensional gas density distribution is derived from the observed X-ray surface brightness profile by inverting Equation~\ref{1.em.x.eq22}. However, if the ICM is clumpy, the gas density inferred from the X-ray surface brightness is overestimated by $\sqrt{C(r)}$ and the gas entropy $S \equiv T/n_e^{2/3}$ is underestimated by $C(r)^{1/3}$. 

Our goal in the X-ray cluster analysis is to recover both the electron density $n_e(r)$ and the scatter in the gas density distribution $\sigma_{n_e,{\rm intr}}$ (and hence the clumping factor $C(r)$) by deprojecting the X-ray surface brightness profile. From the two-dimensional surface brightness map ${\bf \sf S_X}$, we first compute the 1D surface brightness profile $S_X(r)$ (i.e. the surface brightness averaged in circular annuli). We assume that the cluster is spherically symmetric, and it has an onion--like structure with concentric spherical shells, each characterized by uniform gas density and temperature within it. We define a matrix ${\bf \sf V_i^j}$, which contains the effective volumes, i.e. contributions of the volume fraction of the $j$-th spherical shell to the $i$-th annulus. The relation between the surface brightness $S_X$ and the gas density $n_e$ (Equation~\ref{1.em.x.eq22}) can be then expressed using the following matrix formalism:
\begin{equation}
S_X=\frac{1}{4 \pi (1+z)^4} \Lambda(T^*_{\rm proj},Z) \; {\bf \sf V}\#\left({C n_e^2}\right)/{\bf A}+{\bf \sf N},
\label{kk}
\end{equation}
where $T^*_{\rm proj}$ is the observed (projected) temperature, $n_e=<n_e>$ is the average value of the gas density in the $j$-th shell, ${\bf A}$ is the area of the annuli and the operator $\#$ indicates the matrix product (rows by columns). ${\bf \sf N}$ is the noise vector at the $i$-th annulus, $\sigma_{S_X,{\rm tot}}$, representing the observed surface brightness inhomogeneities. $\sigma_{S_X,{\rm tot}}$ is the sum of the intrinsic scatter in the surface brightness ($\sigma_{S_X,{\rm intr}}$) and the Poisson noise ($\sigma_{S_X,{\rm noise}}$):
\begin{equation}
\sigma_{S_X,{\rm tot}}^2=\sigma_{S_X,{\rm intr}}^2+\sigma_{S_X,{\rm noise}}^2 \ .
\label{eqn:mm4dd}
\end{equation}
$\sigma_{S_X,{\rm noise}}$ is simply the square-root of the total (source+background) counts.

From Equation~\ref{eqn:mm4dd} we can infer $\sigma_{S_X,{\rm intr}}$, which contains the information we are looking for (i.e. $\sigma_{n_e,{\rm intr}}$) and it is related to the covariance matrix ${\bf \sf C}_{\rm n_e^2}$ of $n_e^2$ via the following relation:
\begin{equation}
{\bf \sf C}_{n_e^2} = ({\bf \sf V}^{\rm t}{\bf \sf C}_{S_X}^{-1} {\bf \sf V})^{-1} \ ,
\label{eqn:mm4}
\end{equation}
where ${\bf \sf C}_{S_X}$ is the intrinsic scatter diagonal matrix $\left[\sigma_{S_X,{\rm intr}}\right]^{2}$.

In order to recover $\sigma_{S_X,{\rm intr}}$, we re-bin the observed surface brightness map ${\bf \sf S_X}$ such that there are enough ($\gesssim15-20$) counts per pixel to assume Gaussian total errors. We evaluate our ability to significantly measure $\sigma_{S_X,{\rm intr}}$ by using the following F-test: 
\begin{equation}
F=\frac{\sigma_{S_X,{\rm tot}}^2}{\sigma_{S_X,{\rm noise}}^2} \ ,
\label{eqn:mm4ddb}
\end{equation}
which has the F-distribution under the null hypothesis (i.e. the total scatter is consistent with Poisson noise).

\section{X-ray analysis of Abell~133}\label{dataan}
A133 is an X-ray luminous cluster at $z = 0.0566$, which has a cooling flow core, a cD galaxy and a diffuse, filamentary radio source roughly 40 kpc northwest of the cD that has been classified as a radio relic. The cluster shows substructure, which may indicate that it is undergoing a merger, e.g. an irregular morphology in the cluster core. The dominant feature is a tongue-like structure of cool X-ray gas which extends to the northwest with respect to the center, and whose wings are likely due to the passage of a weak shock through the cool core \citep{fujita2002}. Its X-ray spectrum indicates that the emission is thermal, with temperature lower than that of the ambient hot gas. Moreover, the X-ray surface brightness at the position of the radio relic is smaller than the surrounding region except for the tongue, suggesting that the radio plasma has displaced the thermal gas in this region. On larger scales, there is evidence for cluster substructure (see Figure \ref{efffy5n3fr}).

We measured a central cooling time $t_{\rm cool}\simeq 2.4\times 10^9$~yr, which is considerably less than the age of the universe. The cool-core corrected X-ray temperature is $T_{\rm X}=4.39\pm0.04$~keV, and the abundance is $Z=0.62\pm0.03\, Z_{\odot}$. Like other cool-core clusters, A133 shows a strong spike in the X-ray surface brightness profile and a drop in the temperature with $T\sim 3$~keV in the central region. 

Description of the X-ray analysis methodology can be found in \cite{morandi2013}. Here we briefly summarize the most relevant aspects of our data reduction and analysis of Abell~133.

\subsection{X-ray data reduction}\label{laoa}
All data were reprocessed from the level 1 event files using the \textit{CIAO} data analysis package -- version 4.5 -- and the latest calibration database (CALDB 4.5.6) distributed by the {\it Chandra} X-ray Observatory Center. We analyzed 33 datasets retrieved from the NASA HEASARC archive with a total exposure time of approximately 2.4~Msec. One observation (ID 2203) was carried out using the ACIS--S CCD imaging spectrometer and it is telemetered in Faint mode; the remaining observations by using the ACIS--I CCD and they are telemetered in Very Faint mode (see Table \ref{tabdon}). 

\begin{table}
\begin{center}
\caption{Particulars of the A133 observations, including the observation ID, the CCD Imaging Spectrometer, the observation mode and the effective exposure time (ks).}
\begin{tabular}{l@{\hspace{1.8em}} c@{\hspace{1.8em}} r@{\hspace{1.8em}} c@{\hspace{1.8em}}  }
\hline
ID & CCD & MODE & Effective Exp.   \\
   &     &      &  {(ks)\quad } \\
\hline \\
13442 & ACIS--I & VFAINT &  176  \\ 
13443 & ACIS--I & VFAINT &   69  \\ 
13444 & ACIS--I & VFAINT &   38  \\ 
13445 & ACIS--I & VFAINT &   65  \\ 
13446 & ACIS--I & VFAINT &   58  \\ 
13447 & ACIS--I & VFAINT &   69  \\ 
13448 & ACIS--I & VFAINT &  146  \\ 
13449 & ACIS--I & VFAINT &   67  \\ 
13450 & ACIS--I & VFAINT &  108  \\ 
13451 & ACIS--I & VFAINT &   70  \\ 
13452 & ACIS--I & VFAINT &  136  \\ 
13453 & ACIS--I & VFAINT &   69  \\ 
13454 & ACIS--I & VFAINT &   92  \\ 
13455 & ACIS--I & VFAINT &   69  \\ 
13456 & ACIS--I & VFAINT &  135  \\ 
13457 & ACIS--I & VFAINT &   69  \\ 
14333 & ACIS--I & VFAINT &  135  \\ 
14338 & ACIS--I & VFAINT &  117  \\ 
14343 & ACIS--I & VFAINT &   35  \\ 
14345 & ACIS--I & VFAINT &   34  \\ 
14346 & ACIS--I & VFAINT &   81  \\ 
14347 & ACIS--I & VFAINT &   68  \\ 
14354 & ACIS--I & VFAINT &   39  \\ 
13391 & ACIS--I & VFAINT &   46  \\ 
13392 & ACIS--I & VFAINT &   49  \\ 
13518 & ACIS--I & VFAINT &   50  \\ 
2203  & ACIS--S &  FAINT &   32  \\ 
3183  & ACIS--I & VFAINT &   45  \\ 
3710  & ACIS--I & VFAINT &   44  \\ 
12177 & ACIS--I & VFAINT &   49  \\
12178 & ACIS--I & VFAINT &   47  \\
12179 & ACIS--I & VFAINT &   51  \\      
9897  & ACIS--I & VFAINT &   69  \\
\hline        
\label{enf}   
\end{tabular} 
\label{tabdon}
\end{center}
\end{table}   

\begin{figure*}
\begin{center}
\psfig{figure=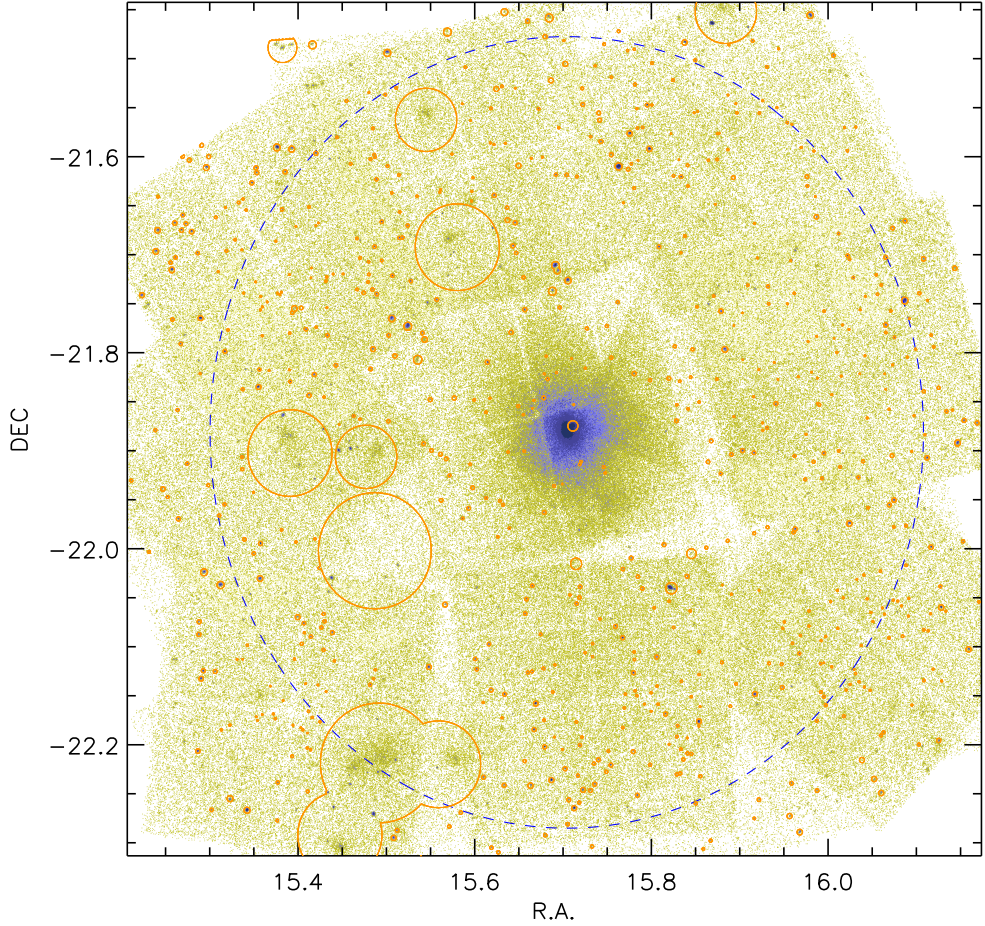,width=0.8\textwidth}
\caption[]{\emph{Chandra} mosaic image of A133. The pixel size of the image is about 1~kpc. The blue circle indicates $R_{200}$ ($R_{200}=1596\pm44$~kpc), while the orange circles indicate the masked point sources and substructures.}
\label{efffy5n3fr}
\end{center}
\end{figure*}                           
We reprocessed the level-1 event files to include the appropriate gain maps and calibration products. We used the \texttt{acis\_process\_events} tool to check for the presence of cosmic-ray background events, correct for spatial gain variations due to charge transfer inefficiency and re-compute the event grades. Then we filtered the data to include the standard events grades 0, 2, 3, 4 and 6 only, and therefore we filtered for the Good Time Intervals (GTIs) supplied, which are contained in the {\tt flt1.fits} file. We then used the tool {\tt dmextract} to create the light curve of the background. Indeed a careful screening of the background light curve is necessary for a correct background subtraction and to discard contaminating flare events. In order to clean the datasets of periods of anomalous background rates, we used the {\tt deflare} script, so as to filter out the times where the background count rate exceed $\pm 3\sigma$ of the mean value. Finally, we filtered the ACIS event files on energy selecting the range 0.3-12 keV and on CCDs, so as to obtain a level-2 event file. 

The bright point sources ($\sim 700$) were identified and masked out using the script \texttt{vtpdetect}, which provides candidate point sources, and the result was then checked through visual inspection. In addition, we masked out a few additional substructures on large-scale (see Figure \ref{efffy5n3fr}) and near the core. Candidates substructures are proposed via both wavelet analysis and visual inspection on brightness image, enhanced with an unsharp mask and filtered Radon back-projection (\S \ref{concl343}). We then calculated the brightness of clumps after subtracting the azimuthally-averaged X-ray brightness. If the clump brightness is larger than five times the brightness standard deviation in the neighboring annulus (\S \ref{phys118d}), the substructure is masked out, as it does not reflects the average properties of the cluster.

We then produced X-ray images and performed a spectral analysis. Our goal is indeed to measure the gas density and clumping factor profile in a non-parametric way from the surface brightness, and to infer the projected temperature profile by analyzing the spectral data.

The X-ray images were extracted from the level-2 event files in the energy range $0.5-2.0$ keV, then corrected by the exposure map to remove the vignetting effects. We created an exposure-corrected image from a set of observations using the \texttt{merge\_obs} to combine the ACIS--I observations (see Figure \ref{efffy5n3fr}). All maps were checked by visual inspection at each stage of the process.

We then determined the centroid ($x_{\rm c},y_{\rm c}$) of the surface brightness image by locating the position where the derivatives of the surface brightness variation along two orthogonal (e.g., X and Y) directions become zero, which is usually a more robust determination than a center of mass or fitting a 2D Gaussian if the wings in one direction are affected by the presence of neighboring substructures. The cluster center that we selected ($01^{\rm h}02^{\rm m}41\fs8$; $-21\degr52\arcmin52.7\arcsec$, J2000) is the centroid of the surface brightness image. We found that this center is slightly offset from the position of the cD galaxy ($01^{\rm h}02^{\rm m}41\fs7$; $-21\degr52\arcmin56\arcsec$, J2000) \citep[see, also,][]{randall2010}: the shift between them is $\simeq 4$ arcsec. This offset suggests that the core of A133 has a disturbed morphology in the inner volumes, as previously pointed out. 

We also measured the flattening and orientation of the X-ray surface brightness. We computed the moments of the surface brightness within a circular region of radius 800 kpc centered on the centroid of the X-ray image (see \cite{morandi2010a} for further details on this method). We rebinned the mosaic image by factor of 16 and we excluded in the analysis the regions previously masked out. The centroid calculated via the surface brightness moments is in agreement with the centroid previously determined by locating the position where the X and Y derivatives of the brightness go to zero, the distance between them $\simeq 7$ arcsec (the uncertainty on this measure is comparable to the rebinning scale applied to the X-ray image). This agreement of the centroids recovered via two independent methods and the smooth X-ray isophotes, suggest that A133 is reasonably relaxed on large-scale out to $\sim 0.5\, R_{200}$, once local small substructures and point sources are masked out. Moreover, the brightness distribution appear to be slightly elongated, with an axial ratio $1.078\pm0.002$ and position angle $19.5\pm0.6$ degrees (measured north through east in celestial coordinates).

The spectral analysis was performed by extracting the source spectra from circular annuli around the centroid of the surface brightness and by using the \emph{CIAO} \texttt{specextract} tool from each observation. The spectral fit was performed by simultaneously fitting an absorbed APEC emission model \citep{foster2012} in the energy range 0.6-7 keV (0.6-5 keV for the outermost annulus only).  We fixed the redshift to the value obtained from optical spectroscopy ($z=0.0566$) and the absorbing equivalent hydrogen column density $N_H$ to the value of the Galactic neutral hydrogen absorption derived from radio data \citep{1990ARA&A..28..215D} ($N_H=0.0157\times 10^{22} {\rm cm}^{-2}$). We also group photons into bins of at least 20 counts per energy channel and applying the $\chi^2$-statistics. We consider three free parameters in the spectral analysis for each annulus: the normalization of the thermal spectrum $K_{\rm m} \propto \int n^2_{\rm e}\, dV$; the emission-weighted temperature $T^*_{\rm proj,m}$ and the metallicity $Z_{\rm m}$ retrieved by employing the solar abundance ratios from \cite{grevesse1998}. We also analyzed the observations individually to check for consistency before analyzing a joint dataset. 

\begin{figure}
\begin{center}
\psfig{figure=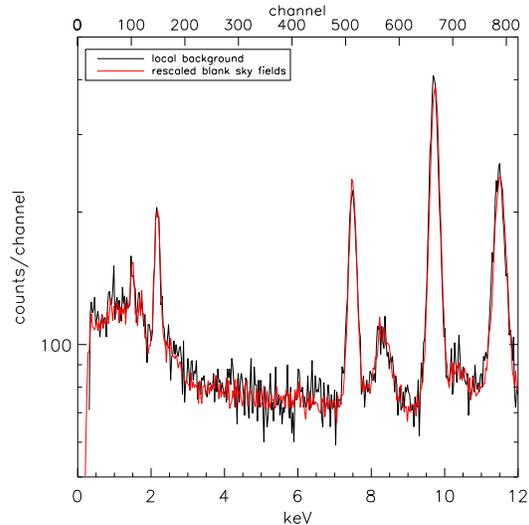,width=0.44\textwidth}
\caption[]{Comparison between the local background and the re-scaled blank sky field spectrum for the observation ID 14338. The two spectra have been extracted in the same area of the CCD free from source emission. We scaled the blank sky spectrum level to the corresponding observational spectrum in the 9.5-12 keV range and to the same integration time. We also rebinned both spectra by a factor of 2, in order to reduce Poisson noise.}
\label{effdddd}
\end{center}
\end{figure}

Diffuse emission from Abell 133 fills the image field-of-view for each observation, so it is difficult to estimate the background from the same dataset. We therefore used the ACIS ``blank-sky" background files appropriate for each observation. We first extract the blank-sky spectra from the blank-field background data provided by the ACIS calibration team in the same chip regions as in the observed cluster spectra. The blank-sky observations underwent a reduction procedure consistent with the one applied to the cluster data, after being reprojected onto the sky according to the observation aspect information by using the {\tt reproject\_events} tool. We then scaled the blank sky spectrum level to the corresponding observational spectrum in the 9.5-12 keV interval, since in this band the \emph{Chandra} effective area is negligible and thus very little cluster emission is expected. One of the advantages of this method is that the derived ARF and RMF will be consistent both for the source and the background spectra. However, the background in the X-ray soft band ($\lesssim 2$~keV) can vary both in time and in space, so it is important to check whether the background derived by the blank-sky datasets is consistent with the real one. In this respect, we verified that for those ACIS--I observations for which we have some areas free from source emission that the two background spectra are very similar (see Figure~\ref{effdddd} and \S \ref{sys453}).

\begin{figure*}
\begin{center}
 \hbox{
\psfig{figure=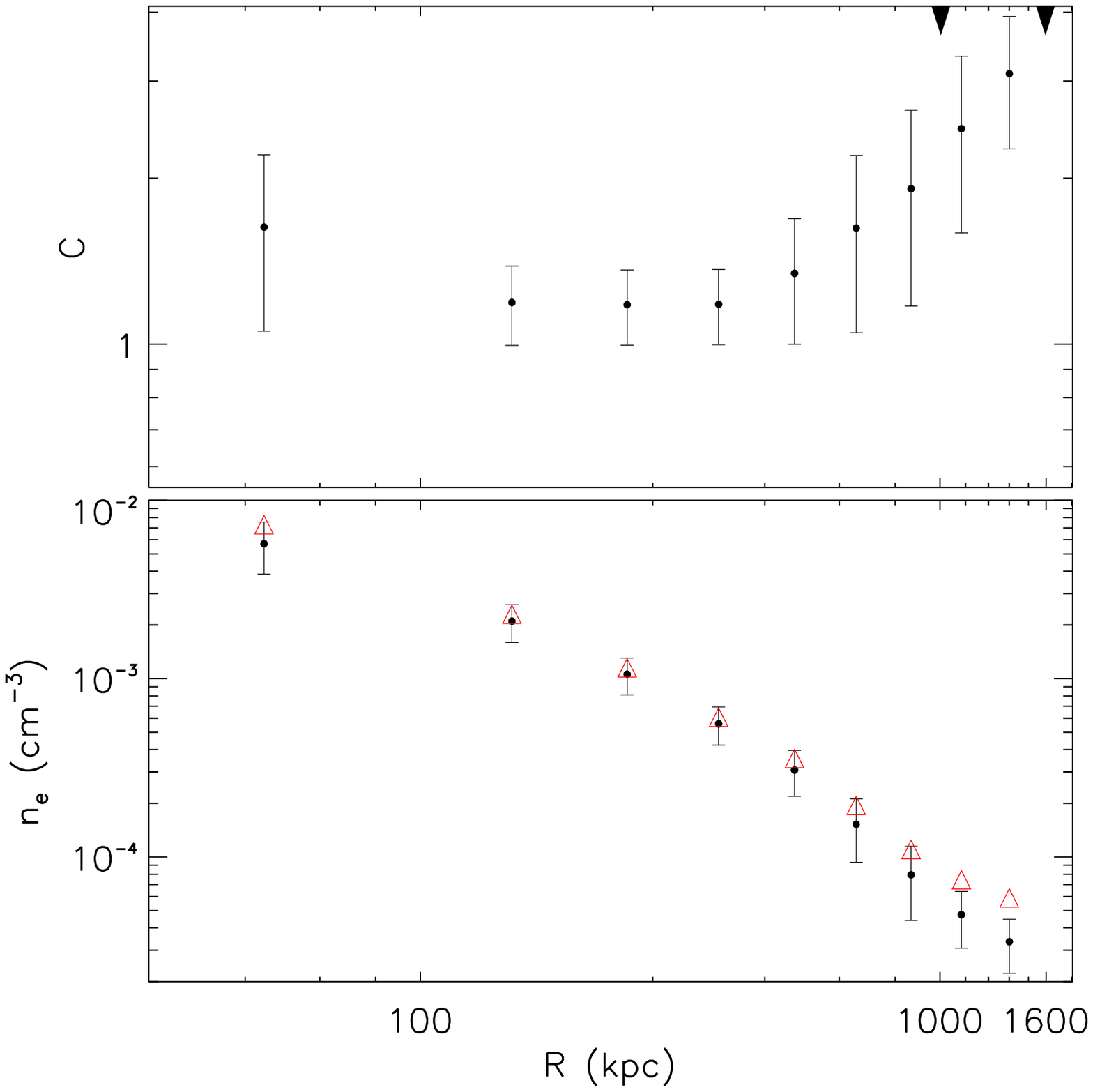,width=0.5\textwidth}
\psfig{figure=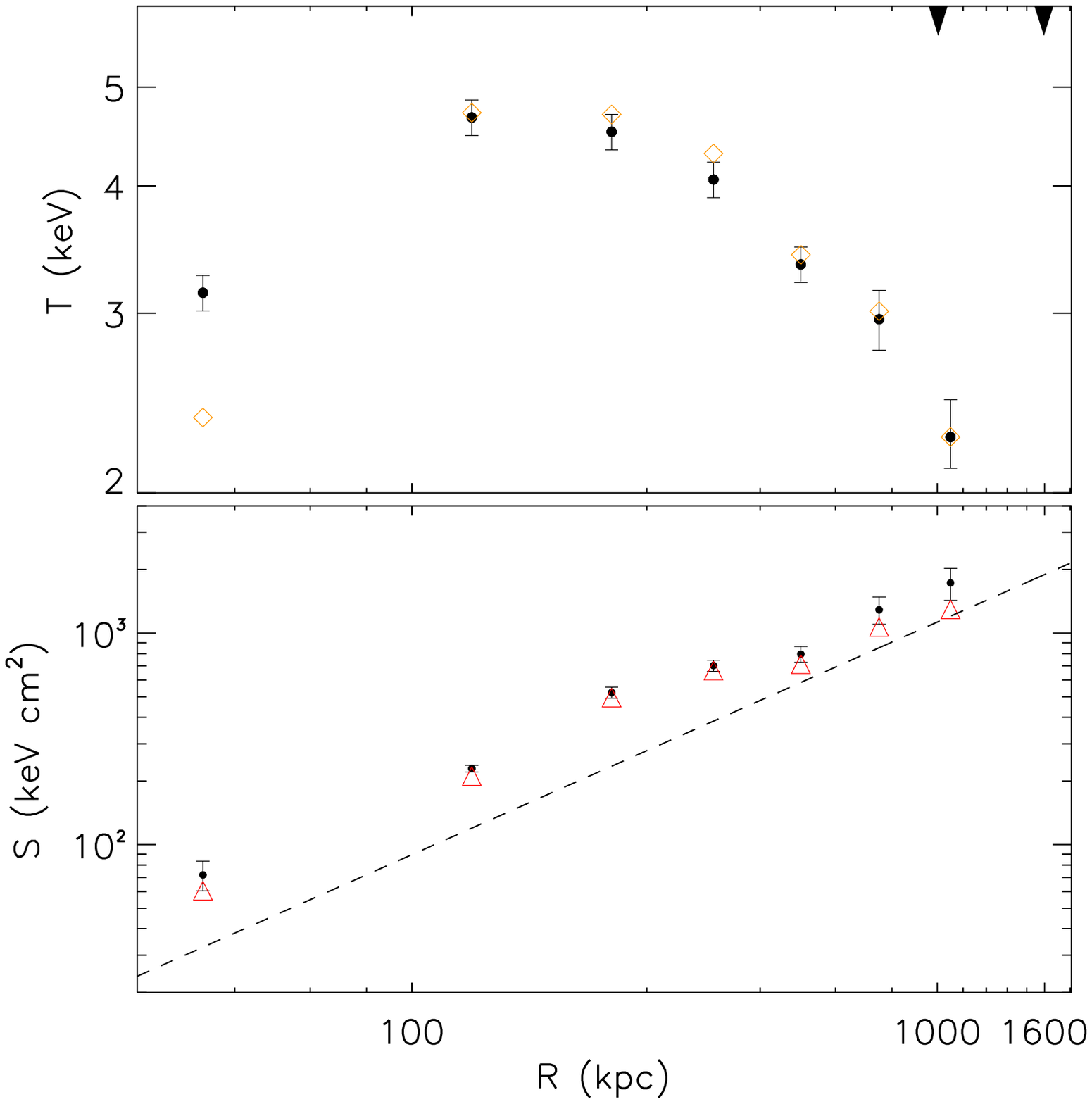,width=0.5\textwidth}
}
\caption[]{\emph{Chandra} X-ray observations of A133. {\it Left panel:} 3D gas clumping factor (upper panel) and gas density (lower panel) profiles. {\it Right panel:} observed and deprojected temperature profile (upper panel) and entropy profile (lower panel). In the bottom panels, we show results with (circles) and without (triangles) the gas clumping factor. In the top-right panel, the solid circles and diamonds show the projected and deprojected temperature profiles respectively. For the deprojected temperature we omitted the errorbars for clarity. For the entropy profile, the dashed line represents the predictions of \cite{Voit2005b} from pure gravitational collapse, where the entropy is defined as $S(r)=S_{200}\,1.32(r/R_{200})^{1.1}$, $S_{200}$ being a characteristic value of the entropy at an overdensity of 200 \citep[see, e.g., Eq.~2 in][]{Voit2005b}. From the left to the right, the arrowhead pointers at top of the upper panels indicate $R_{500}$ and $R_{200}$, respectively.}
\label{entps3xkn3fr}
\end{center}
\end{figure*}

\subsection{Reconstruction of gas clumpiness and density in A133}\label{phys118d}
We re-binned the stacked surface brightness map ${\bf \sf S_X}$ by a binning factor of 32 for each axis, such that that there are more than $15-20$ counts per pixel. We then constructed a set of circular annuli around the centroid of the surface brightness and computed the azimuthally-averaged surface brightness profile $S_X(r)$. We deduced the electron density $n_e$ by deprojecting the surface brightness profile, obtaining radial measurements in spherical shells. Finally, we applied the method described in \S\ref{tecn} on $S_X(r)$ to derive the 3D gas clumping factor profiles. We selected the boundary radius according to the following criteria: the null hypothesis that the total scatter is consistent with Poisson noise is rejected with a probability of 90\% (see Equation~\ref{eqn:mm4ddb}). The boundary radius is set to 1500 kpc in our A133 analysis.

We then inferred $R_{200}$ by  calculating the mass profile under the assumption of hydrostatic equilibrium and spherical symmetry. Specifically, once a Navarro, Frenk and White DM density profile (NFW) is assumed, the theoretical 3D gas temperature is obtained by integration of the hydrostatic equilibrium equation and via the measured gas density. We then conveniently projected the theoretical 3D gas temperature \citep{mazzotta2004}. The comparison of model projected temperature with the measured values allows us to infer the NFW parameters (concentration parameter and scale radius) and thus the mass profile and $R_{200}$ \citep[see][for further details]{morandi2007a}. We found $R_{200}=1596\pm44$~kpc, where the quoted uncertainties reflect both statistical and systematic errors due to background modeling.

We also deprojected the best-fit results of the X-ray spectral analysis using the "onion peeling" method employed by \cite{morandi2007b}. More specifically, we inverted the following equation:
\begin{equation}\label{aa1}
T^*_{\rm proj} = {{\bf \sf V} \# {( {T} w )} } \; / \, {\left({\bf \sf V} \# w \right)},
\end{equation}
where $T$ is the deprojected temperature, $w={{n_e^2 T^{-\alpha}}}$, and $\alpha=0.75$ corrects for the temperature gradient along the line of sight as suggested in \cite{mazzotta2004}. We then recovered the gas entropy $S \equiv T/n_e^{2/3}$.

The left panel of Figure \ref{entps3xkn3fr} shows the derived 3D clumping factor and gas density for A133. The gas clumping factor becomes larger than unity at $r \gesssim R_{500}$, where $R_{500}=1044\pm27$~kpc, reaching $C \approx 2-3$ at $0.9\, R_{200}$. Therefore $R_{500}$ can be regarded as a transition of the smooth state in the virialized region ($\lesssim R_{500}$) to a clumpy intergalactic medium in the infall region ($R_{500}-R_{200}$). Our clumping factor is in good agreement with the predictions of hydrodynamical simulations \citep[e.g.][]{nagai2011}.

In the outer volumes, our gas density profile (without gas clumping factor correction) is in disagreement with the predictions from hydrodynamical numerical simulations including cooling, star formation and supernovae feedback \citep{roncarelli2006} and not corrected for the effect of clumping. We observe a general trend of steepening in the radial profiles of the gas density beyond $0.3\, R_{200}$, with a logarithmic slope of $\sim 2.0$ at $0.9\,R_{200}$ ($\sim 2.6$ with gas clumping factor correction). We see that the predicted density profiles are too steep (the logarithmic slope from simulations is $\sim 2.8$ at $R_{200}$) compared to the data. A possible explanation for this trend is that in the above simulation \cite{roncarelli2006} did not include AGN feedback and preheating. Indeed, recent works \citep[][]{mathews2011} indicate that feedback mechanisms may be responsible for the deficit of baryons in cluster cores, smoothing the accretion pattern and leading to a flatter gas distribution. Therefore, we can see that at this level of precision the effects of additional physics cannot be neglected, even in regions well outside of the cluster core. The agreement with simulation is improved when comparing our findings with the simulations of galaxy clusters formation from \cite{nagai2007a}, which were performed using the ART code \citep{kravtsov2002}. Their clumping-corrected gas density profiles indicate a slope $\sim 2.7$ at $R_{200}$ \citep[see][]{eckert2011}, just slightly steeper than our profile. 

Additionally, \cite{eckert2011} performed a stacking of the density profiles of a sample of clusters observed through \emph{ROSAT} to analyze the outskirts of clusters, although they cannot determine any spectral information (i.e. the gas temperature) from these data. Their average density profile (which is not corrected for clumping factor) steepens beyond $R_{500}$, with a logarithmic slope of $\sim 2-2.3$ in the range $(0.3-1)\, R_{200}$, in agreement with the present work and with the results from previous \emph{ROSAT} observations \citep{vikhlinin1999}.

We point out that the angular resolution of the rebinned image ($\approx 16$~arcsec, which corresponds to the physical scale of about $17$~kpc) sets the smallest scale down to which we can resolve gas clumps. Note also that in \cite{morandi2013} we tested our method on the simulated clusters which resolves dense gas clumps down to about $10$~kpc, demonstrating that we can indeed recover the true clumping factor in the simulations from the rebinned image. However, since our method does not detect the small-scale clumps below $17$~kpc, we emphasize that the clumping factor inferred from our method should be taken as a lower limit of the true clumping factor. As we discussed in \cite{morandi2013}, the clumping factor is reconstructed with a bias of 5-10 percent, primarily due to the combination of substructures and asphericity in the gas distribution.

In the literature, clumpiness has been also constrained via {\emph Suzaku}, where measurements of gas clumping are indirectly inferred from its effects on thermodynamic profiles. For example, gas clumpiness can be indirectly inferred under the assumption that the measured entropy profiles lies above its theoretical predictions only because of gas inhomogeneities \citep[see,e.g.][]{walker2013,urban2013}; or by assuming that gas fraction exceeds the cosmic mean baryon fraction measured from the CMB due to the presence of gas density clumps \citep{simionescu2011}. We observe that there is a good agreement between our clumping factor and the determinations of the gas clumping from {\emph Suzaku} \citep[$C\sim 2$ at $R_{200}$, see][]{walker2013}, which hinge on independent assumptions with respect to the our measurements. We also find a good agreement with the findings from hydrodynamical numerical simulations \citep{nagai2011,vazza2013,roncarelli2013}. This suggests that our clumping factor determination is not significantly biased by the adopted resolution in the rebinned image.

The right panel of Figure \ref{entps3xkn3fr} shows the temperature profile of A133, which shows a drop by roughly a factor of 2 from the peak temperature to $r\approx 0.7\, R_{200}$. Next we focus on the entropy, since this is a powerful tool to trace the thermal history of the ICM. In the central regions, galaxy clusters are known to exhibit an excess of entropy with respect to the prediction from pure gravitational collapse. This excess in the entropy (labeled the entropy ``floor'' or ``ramp'') calls for some energetic mechanism, in addition to gravity, such as (pre)-heating and cooling \citep{Borgani2005,bryan2000,morandi2007b}. Somehow these non-gravitational processes intervene to break the expected self-similarity of the IC gas in the innermost regions, this effect being stronger in groups than massive clusters. Nevertheless, in the outer volumes simple theoretical models predict that the entropy $S$ should be self-similar and behave as a power-law with radius. Models of shock dominated spherical collapse show that matter is shock heated as it falls into clusters under the pull of gravity, with a slope of $\sim$1.1 \citep{tozzi2001,Voit2005b}. 

With respect to these theoretical predictions, the results presented here point to an entropy excess in the central regions, which extends out to large radii. When not corrected for the clumping factor, the entropy profile tends to show a flatter profile, in agreement with most of the results obtained from {\em Suzaku} \citep{george2009,simionescu2011,walker2013}. The entropy profile match the self-similar expectation at $\sim 0.7\, R_{200}$, in roughly agreement with the findings of \cite{eckert2013}, who also argued for an entropy excess significantly beyond the cluster core. These results highlight the importance of correcting for the clumping factor while comparing with theoretical predictions and in order to use galaxy clusters as high-precision cosmological probes.

\begin{figure}
\begin{center}
\psfig{figure=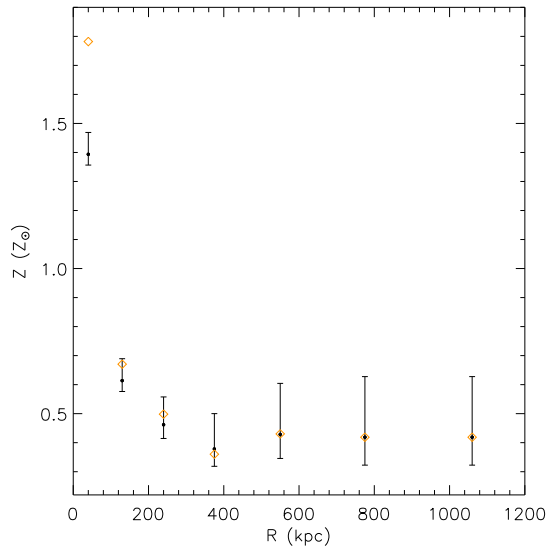,width=0.44\textwidth}
\caption[]{Metallicity profiles for \emph{Chandra} X-ray observations of A133 (in units of $Z_\odot$,  which is the solar abundance of iron). The solid circles and diamonds show the projected and deprojected metallicity profiles, respectively.}
\label{eerggdg33d}
\end{center}
\end{figure}

Finally, we discuss the metal abundance in the ICM. \emph{Chandra} data sets measure the metallicity structure of the intra-cluster gas with high precision and spatial resolution roughly out to $0.7\,R_{200}$. We observe a centrally peaked metal abundance, which is likely due to large contribution of SNe Ia products from the cD galaxy. Moreover, our results show that the cluster outskirts are also substantially metal-enriched, to a level amounting to approximately one third of the Solar metallicity (see Figure \ref{eerggdg33d}). The ICM metal content is a key observable to constrain the cumulative past star formation history in galaxy clusters and to study the enrichment processes. While the production of metals is linked to processes of star formation, its radial profile is determined by different physical processes, such ram-pressure stripping, galactic winds powered by supernovae and AGN activity, merger mechanism \citep{gnedin1998}. These results therefore can provide an anchor for numerical simulations of ICM physics in the outskirts, constraining the metal enrichment processes of the ICM.

The large metal content out to the outer volumes is a proxy of feedback processes due to star formation, which release energy into the ICM and break its self-similarity. This is in agreement with the entropy excess with respect to models of pure gravitational collapse that we previously discussed.

\subsection{Analysis of the systematics}\label{sys453}
In this Section we discuss how different sources of systematics can affect our measurements, in particular with respect to the background modeling. Indeed the ratio of source to background flux is $\sim$ 8 percent in the outermost annulus, hence these systematics (in particular the modeling of the particle background and unresolved point sources) can bias the measurement of the physical parameters in the outer volumes due to the low signal-to-noise. 

A key aspect is the modeling of the particle background, which dominates the X-ray spectrum at energies $\gesssim 2$~keV. In this respect, one source of uncertainty is the amount by which the blank sky fields are to be rescaled to match the cluster count rate at high energy (9.5-12 keV). The relative error in the rescaling of the background to match the cluster count rate at high energy is $\lesssim 1$ percent, as determined by the Poisson error in the photon counts at high energy. 

Another source of biases is represented by possible variations in the background spectrum. \cite{hickox2006} has shown that the spectral distribution of the particle background is remarkably stable, even in the presence of changes in the overall flux, and that the ratio of soft-to-hard (2-7 keV to 9.5-12 keV) count rates remains constant to within $\lesssim 2$ percent. We therefore apply a systematic error of 2 percent in the background flux, to account for possible uncertainties in the background spectrum. We used the task {\it grppha} to set the fractional systematic errors associated with each channel in the PHA file for the spectral analysis.

Additional uncertainties are associated with the background in the X-ray soft band ($\lesssim 2$~keV), in particular with the modeling of the Galactic foreground components. The background in the X-ray soft band can indeed vary both in time and in space, so it is important to check whether the background derived by the blank-sky datasets is consistent with the real one. In this respect, we verified that for those ACIS--I observations for which we have some areas free from source emission that the two background spectra are consistent (the reduced $\chi^2$ is smaller than 1) within the statistical errors (see Figure~\ref{effdddd}). We account for the uncertainties in assuming that the background matches the cluster count rate in the soft band as Poisson errors in the photon counts in the range 0.5-2 keV.

Finally, we estimated the error due to unresolved point sources. As the cosmic X-ray background (CXB) consists of unresolved point sources, which are not uniformly distributed across the sky, the CXB level deviates from the mean when analyzing regions of finite size due to varying numbers of unresolved point sources. Following \cite{walker2013,urban2013}, the expected deviation from the average value for a given observed solid angle $\Omega$ resolved to a threshold flux $S_{\rm thres}$ is:
\begin{eqnarray}
\sigma^2_{\rm CXB} = (1/\Omega) \int_{0}^{S_{\rm thres}} \Big(\frac{dN}{dS}\Big)\times S^2 ~ dS
\label{flux56}
\end{eqnarray} 
where ${dN}/{dS}$ is the cumulative flux distribution of point sources as employed by \cite{moretti2003}. The threshold flux $S_{\rm thres}$ has been calculated by determination of the local flux limit to which we can robustly identify a point source, commonly known as the sensitivity map \citep[see, e.g.,][for further details on the method]{ehlert2013}. Our joint Chandra observations of the outskirts allow the CXB to be resolved to a threshold flux $S_{\rm thres} \sim1.3\times10^{-15}$ erg cm$^{-2}$ s$^{-1}$ deg$^{-2}$ in the soft band. Solving Equation \ref{flux56}, this translates into an error up to 7\% on the surface brightness in the outer volumes, while it becomes negligible in the inner volumes. 

In this respect, concerning the surface brightness analysis, we produced Montecarlo (MC) randomizations of the background including both statistical and systematic uncertainties. In particular, uncertainties due to unresolved point sources (Equation \ref{flux56}) and to background normalization were added in quadrature to the statistical errors. We then performed the deprojection for each MC realization, in order to correctly propagate these biases on the gas density and clumping factor. We verified that these systematic uncertainties have a small impact on the recovery of the 3D gas density and hence gas clumping factor profiles. For example, the gas density is affected by a bias up to 25 percent (in the outermost annulus): although not negligible, these biases lie within the statistical errors.

Concerning biases on the temperature, we found that the spectra of the resolved point sources extracted from the data are in good agreement with a power law of index $\sim 1.4$ \citep[see, also][]{moretti2003,walker2013}. We then used this spectrum as the input spectral model in order to account for the uncertainty in the spectral fit due to unresolved point sources. Specifically, for each annulus we performed a Poisson randomization of unresolved point sources according to the distribution employed by \cite{moretti2003}. We then simulated input spectra of point sources, and we added them in the background model for a given energy bin. In the background model we includes statistical and systematic errors, and a MC randomization of the background scale factor according to the Poisson error in the photon counts at high energy. We then applied the spectral fit for each MC iteration. We find that the uncertainties due to background systematics are smaller than the statistical error bars for all of our temperature measurements.

\section{A snapshot of the Large Scale Structure Formation Scenario}\label{concl343}
In this section we investigate whether we can unveil from our data a record of the large-scale structures formation scenario, e.g. in the brightness morphology, gas density and clumpiness distribution.

We applied a filter in order to amplify the high-frequency components of the brightness image of A133 and detect subtle features otherwise overwhelmed by noise. This filter applies (in the following order): i) an unsharp mask technique, which subtracts a symmetric analytical model of the surface brightness from the original image; ii) a filtered Radon back-projection. 

Concerning the first point, we consider the following symmetric analytical model as an approximation of the cluster global surface brightness distribution and we subtract it from the original image:
\begin{eqnarray}
n_e(r) = {n_0\; (r/r_c)^{-\delta}}
{(1+r^2/r_c^2)^{-3/2 \, \varepsilon+\delta/2}}
\label{eq:density:model}
\end{eqnarray}
with parameters ($n_0,r_c,\varepsilon,\delta$). This filter enhances azimuthal variation of the brightness with respect to a symmetric surface brightness profile.

Concerning the second point, we used the Radon transform to detect features within an image \citep{toft2009}. Given an image $I$, the Radon transform is defined as:
\begin{eqnarray}
R(\theta,\rho)=\int_{-\infty}^{\infty} I(\rho\cos{\theta}-t\, \sin{\theta},\rho\sin{\theta}+t\, \cos{\theta})\, dt
\label{eq:density:model2}
\end{eqnarray}
This equation describes the integral along a line $t$ through the image $I$, where $\rho$, $\theta$ are the polar coordinates. Note that the argument of the integral in Equation \ref{eq:density:model2} represents the parameterization of any straight line $L$ (at a distance $\rho$ from the origin and with an angle $\theta$ with the $x$-axis) with respect to the arc length $t$. It follows that the Radon transforms lines through an image to points in the Radon domain (called ray-sums); conversely, point-like sources in an image appears in the Radon domain as a number of sine waves with different amplitudes and phases. 

An image can be reconstructed from its ray-sums using the backprojection operator: 
\begin{eqnarray}
I(x,y)=\int_{0}^{\theta} R(\theta,x\,\cos{\theta}+y\, \sin{\theta})\, d\theta
\label{eq:density:model2nn}
\end{eqnarray}
Therefore the general idea is that features within a two-dimensional image, e.g. filamentary structures, roughly correspond to points in the Radon domain, which can be filtered and isolated. A filtered Radon back-projection would then enhance the image features. 

\begin{figure*}
\begin{center}
\psfig{figure=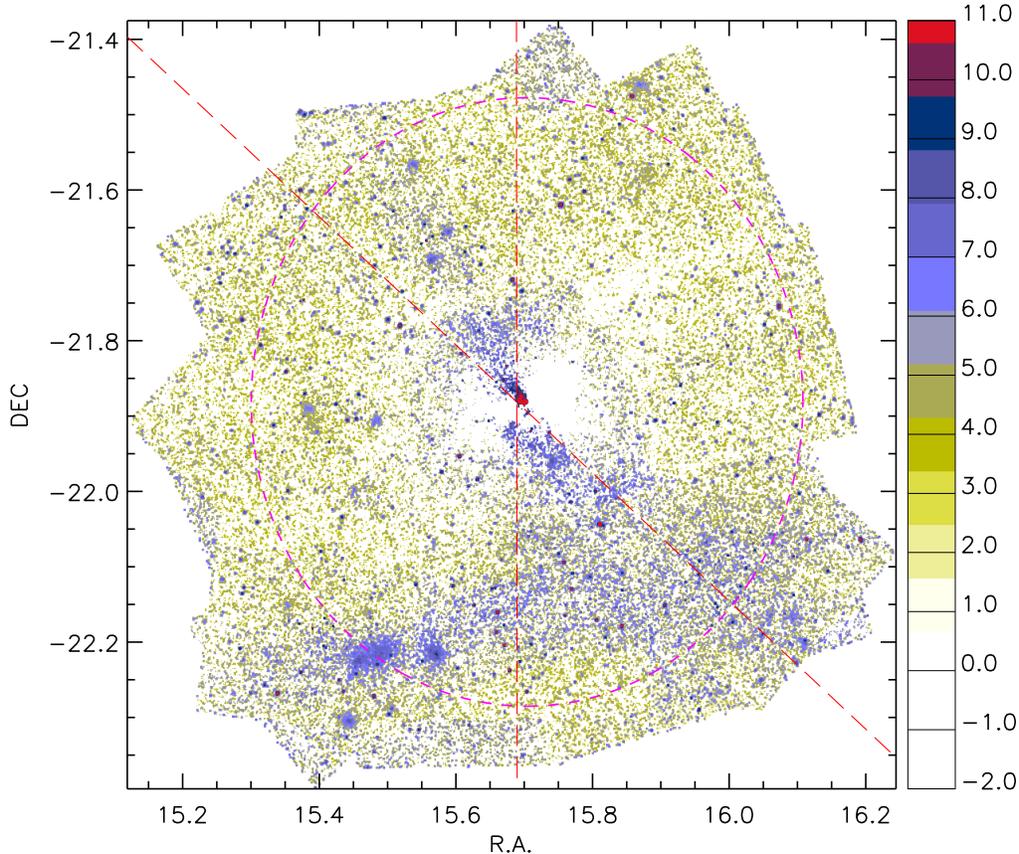,width=0.8\textwidth}
\caption[]{\emph{Chandra} mosaic image of A133, enhanced with an unsharp mask and filtered Radon back-projection. The pixel size of the image is about 4~kpc. The red circle indicates $R_{200}$, while the long-dashed lines define a sector with position angle in the range $0-45$ degrees and encompassing the funnel-like structures in the inner volumes. In the intensity scale, the positive (negative) values are indicative of an excess (defect) of brightness with respect to the cluster global surface brightness distribution.}
\label{effddfr}
\end{center}
\end{figure*}

We applied the filters previously described in the brightness image extracted in the energy range $0.5-2$ keV. For the filtered Radon back-projection, we filter out the frequencies below $0.7$ the maximum ray-sum. The enhanced image is shown in Figure \ref{effddfr}.

This enhanced and filtered image shows a complex, non-spherical morphology of the ICM, with a mild excess of brightness in the south region and along two funnel-like structures with position angle $\sim 20$ degrees (measured north through east in celestial coordinates), and which are approximately symmetric with respect to the center. The south-east region is characterized by the presence of substructures. 

In order to gauge the bias implicit in our spherical modeling, we repeat the X-ray analysis (\S \ref{phys118d}) extracting surface brightness profiles for the $0.5-2$ keV band for two sectors shown in Figure \ref{effddfr}. The first sector has position angle in the range $0-45$ degrees, i.e. encompassing the previous funnel-like structures, while the second characterizes the remaining volumes of the cluster. With respect to the azimuthally-averaged density profile, the first sector shows a density higher of $4-40$ percent increasing from the center towards $R_{200}$, while the second a deficit of $3-20$ percent. Although not negligible, these biases are within the statistical errors. The reconstructed clumpiness profiles from the two sectors agree within $\sim10\%$ with the azimuthally-averaged clumping factor profile, and they are in agreement with the biases estimated in our previous work via hydrodynamical numerical simulations \citep{morandi2013}.

N-body+hydrodynamical simulations envisage a picture where galaxy clusters live at the intersection of a thread-like structure called "cosmic web". The infall of material into the most massive dark matter haloes is not spherical but it is expected to be preferentially funneled through the cosmic filaments where the haloes are embedded. The cluster mass haloes would indeed acquire most of their mass from major mergers along the filaments, then relaxing from this chaotic initial state to a quasi-equilibrium via violent relaxation \citep{limousin2013}. Although A133 appears to be reasonably relaxed and virialized within $R_{500}$, as previously discussed (\S \ref{phys118d}), the enhanced and filtered brightness image keeps a 'fossil record' that can be used to unearth the ongoing large-scale structures formation scenario. In particular, the previous funnel-like structures would likely track the large-scale filament where the cluster is embedded. The gas density along the directions of filaments where the cluster accretes clumpy and diffuse materials would then be enhanced and flattened, with subclumps (in particular in the south-east region) falling into the DM potential well under the pull of gravity. Departures from virialization, clumpy ICM, bulk flows and complex accretion patterns appear to be more pronounced in the outer volumes. This strengthens the picture that we are witnessing the formation of A133 as it happens, with a 'melting pot' in the virialization region, where the surrounding filamentary structure and substructures are infalling into the cluster gravitational potential and converting most of kinetic energy of the fall into the thermal energy of the ICM. 

We also point out that there is a substantial alignment between the major axis of the brightness distribution (with position angle of the major axis $19.5\pm0.6$ degrees, see \S\ref{laoa}) and the large-scale filament. This alignment is expected since the accretion happens preferentially along the cosmic filaments \citep{brunino2007}.

\section{Conclusions}\label{conclusion33}
In this paper, we have presented our analysis of \emph{Chandra} observations of A133, focusing on the clumping factor and gas density profile of the gas in cluster outskirts. \emph{Chandra}'s superior angular resolution enables robust identification and removal of point sources from the X-ray images, while minimizing the contribution from the bright cluster core to the emission in the outer volumes. We find that the gas clumping factor increases with radius and reaches $2-3$ at $r\gesssim R_{500}$. We then compared our observational results with numerical simulations, and we find a good agreement.

We observe a general trend of steepening in the radial profiles of the gas density beyond $0.3\, R_{200}$, with a logarithmic slope of $\sim 2.0$ at $0.9\,R_{200}$ ($\sim 2.6$ with gas clumping factor correction). The observed density profiles appear to be flatter compared to simulations, but in agreement with previous observational findings. This suggests that an improved physical treatment in the ICM might be needed in hydrodynamical simulations to match the observations.

In addition, we observe that the observed temperature profile decreases steadily with radius toward the outskirts of A133, while the entropy profile increases monotonically with radius $r\gesssim R_{500}$. With respect to theoretical predictions from pure gravitational collapse, the results presented here point to an entropy excess in the central regions, which extends out to large radii. 

The results in the present paper suggest that gas inhomogeneities should be treated properly when interpreting X-ray measurements. In this perspective SZ data could be valuable to obtain (in combination with X-ray) an independent measurement of gas clumping, given the different dependences of X-ray/SZ on the electron density, providing further constraints on physical parameters and critical insights to our understanding of the cluster physical properties out to large radii.

We finally discuss how the brightness distribution keeps a record of the large-scale structures formation scenario, providing a snapshot of the 'melting pot' in the virialization region.

\section*{acknowledgements}
This work was supported in part by the US Department of Energy through Grant DE-FG02-91ER40681 and Purdue University. We are indebted to Joseph Richards and Franco Vazza for valuable comments.

% \bibliographystyle{mn2e}
% \bibliography{master}

\newcommand{\noopsort}[1]{}

\end{document}